\documentclass[12pt]{iopart}

\usepackage{graphics}
\usepackage{subfigure}

\def\Ry{{\,\rm Ry}}

\begin{document}

\title{Accurate PAW datasets for BaFe$_2$As$_2$}

\author{Chao Cao}
 \ead{cao@qtp.ufl.edu}
\author{Yu-ning Wu}
\author{Rashid Hamdan}
\author{Yun-Peng Wang}
\author{Hai-Ping Cheng}%
 \ead{cheng@qtp.ufl.edu}
 \address{Department of Physics and Quantum Theory Project, University of Florida, Gainesville, FL 32611, U.S.A.}

\date{\today}

\begin{abstract}
By carefully choosing parameters and including more semi-core orbitals as valence electrons, we have constructed a high quality projected augmented wave (PAW) dataset that yields results comparable to existing full-potential linearized augmented plane-wave calculations. The dataset was then applied to BaFe$_2$As$_2$ to study the effects of different levels of structure optimization, as well as different choices of exchange-correlation functionals. It is found that the LDA exchange-correlation functional fails to find the correct SDW-AFM ground state under full optimization, while PBE exchange-correlation functional obtains the correct state but significantly overestimates the magnetism. The electronic structure of the SDW-AFM state is not very sensitive to structure optimizations with the PBE exchange-correlation functional because the position of the As atoms are preserved under optimizations. We further investigated the Ba atom diffusion process on the BaFe$_2$As$_2$ surface using the nudged elastic bands (NEB) method. The Ba atom was found to be stable above the center of the squares formed by the surface As atoms, and a diffusion barrier of 1.2 eV was found. Our simulated STM image suggests an ordered surface Ba atom structure, in agreement with Ref. \cite{PRB_Ba122_Co_surface,arxiv.1009.3493}.
\end{abstract}


\maketitle

\section{\label{sec:level1}Introduction}
The recent discovery of iron pnictide superconductor parental materials {\it Ln}FeAsO (1111) \cite{la1111_discover} ({\it Ln}=La, Ce, Pm, ...), {\it Ae}Fe$_2$As$_2$ (122) \cite{ba122_discover} ({\it Ae}=Ca, Sr, Ba, ...), and {\it A}FeAs (111) \cite{Li111_discover_1,Li111_discover_2} ({\it A}=Li, Na, ...) has stimulated tremendous interest and attention starting about a year ago. Experimentally, the ground states of these parental materials are orthorhombic lattice with a spin-density-wave (SDW) anti-ferromagnetic (AFM) ordering. At a certain temperature $T_c$, the parental materials of these compounds exhibit a phase transition from orthorhombic lattice to tetragonal lattice, which is accompanied by the loss of SDW-AFM \cite{Nature.453.899,PhysRevB.78.020503,PhysRevB.78.094517}. This phenomenon was regarded as a key issue and was thought to be in close relationship with the superconductivity behavior. From the very beginning, numerous efforts have been made in order to determine the structure and electronic and magnetic properties of these materials from first principles \cite{singh_prl_100_237003,ccao_1,yildirim_1,yildirim_2}. Both the local density approximation (LDA) and generalized gradient approximation (GGA) to the exchange-correlation functional have been employed, while for electron-ion interactions, various methods including ultrasoft pseudopotential \cite{vanderbilt_uspp,ccao_1,yildirim_2} (USPP), projected augmented wave \cite{bloch_paw} (PAW), and full-potential linearized augmented plane-wave \cite{flapw,singh_prl_100_237003,yildirim_1} (FLAPW) were used. In contrast to most materials, the results of these density functional calculations seem to be extremely sensitive to the choice of exchange-correlation functionals, the optimization method, and the electron-ion interaction models; thus these results scatter over a wide range. Previous studies showed that both USPP and PAW methods tend to overestimate the local magnetic moment on individual Fe atoms in the SDW-AFM state (2.3$\mu_B$ USPP \cite{ccao_1}, 1.5$\mu_B$ FLAPW \cite{PhysRevB.78.020503}, 0.7$\mu_B$ Experiment) as well as the relative energy differences between different magnetic states (14 meV/Fe \cite{ccao_1} USPP, 22 meV/Fe FLAPW\cite{PhysRevB.78.020503}) \cite{ccao_1,yildirim_1,yildirim_2}. Mazin {\it et al.} have performed a detailed and systematic analysis on calculations of the 1111 systems \cite{PhysRevB.78.085104} and concluded that unless a USPP or PAW potential was very carefully constructed, one should always use FLAPW method since it employ fewest approximations. The previous inconsistency or failure of density-functional methods are mostly due to defective USPPs or PAWs. While FLAPW is generally accepted as the most accurate density-functional method, it is computationally much more expensive than both USPP and PAW methods. Therefore, the structure optimizations were not performed in most FLAPW calculations, although Mazin {\it et al.} suggested a full structure optimization using GGA and then electronic structure analysis using LDA functional.

In this paper, we report our efforts to generate a reliable PAW dataset for these materials, using BaFe$_2$As$_2$ as an example. The PAW dataset was constructed for the open-source community density functional code PWSCF\cite{PWSCF}, which can be used to generate the tight-binding Hamiltonian or even down-fold multi-band structures. The tight-binding or down-folded Hamiltonian has been proven to be essential for study of cuprates, pnictides and other unconventional superconductors. We compare all of our results with FLAPW results in detail whenever feasible, and we follow the Mazin {\it et al.} procedure to study the influence due to full relaxation of the crystal structure as well as different exchange-correlation functionals on the calculated results. Finally, we applied the newly constructed PAW datasets to study the structure of BaFe$_2$As$_2$ surface. 

The rest of the paper is organized as follows: in the next section, we briefly present the method and calculation details, especially the parameters we employed to generate the PAW datasets used in this paper; then in section \ref{sec:esp} we compare our PAW results to existing USPP/FLAPW results, together with a detailed discussion about the choice of appropriate exchange-correlation functional and structural optimization level for the iron-pnictides; in section \ref{sec:dfp}, we present the surface properties of BaFe$_2$As$_2$, especially the STM image and the Ba positions on the (001) surface; finally, we summarize this paper and draw conclusions.

\section{Method and Calculation Details}
In principle, one can approach all-electron accuracy with pseudopotential or PAW methods by including more semi-core orbitals into valence orbitals, but in practice it is prohibitively difficult to do so for the ultrasoft pseudopotential method. However, this process is practical with Bl\"{o}ch's PAW method \cite{bloch_paw}. For this study, we follow the slightly modified recipe by Kresse {\it et al.} \cite{vasp_paw}, as implemented in the PWSCF package \cite{PWSCF}. The valence orbitals (including the semi-core orbitals) included in the datasets are listed in TABLE \ref{TAB_semicore}, compared with existing USPPs. For each orbital, two projectors were used to generate the PAW dataset. The generated dataset was carefully tested with logarithmic derivatives, ionization potentials, and simple crystal structures. The logarithmic derivative energy range considered was taken to be from $-7.0 \Ry $ to $ 3.0 \Ry $, and orbital angular momentum $l$ from 0 to 4 ($s$, $p$, $d$, $f$, $g$). These tests ensured the quality of the PAW dataset. Throughout this paper, we used a cutoff energy of $ 40 \Ry $ for plane-waves in both PAW and USPP calculations to ensure convergence, and the irreducible Brillouin zones were sampled with the Monkhorst-Pack scheme \cite{mp_kpoints}, $8\times8\times8$ for the body-centered tetragonal (bct) unit cell (non-magnetic NM/checker-board antiferromagnetic CB-AFM states), and $6\times6\times8$ for the base-centered orthorhombic (bco) unit cell (spin-density wave antiferromagnetic SDW-AFM state). Both LDA and Perdew-Burke-Ernzerhof (PBE) parameterizations \cite{PBE_1} to GGA were used.

\begin{table}
  \caption{parameters used to generate the PAW datasets in this paper. The value of $r_{\mathrm{cut}}$ is in unit of bohr, the number in the parentheses indicates the value used for the orbitals with angular momentum represented by the superscript. The ultrasoft pseudopotentials (USPP) are obtained from the PWSCF website.}
  \label{TAB_semicore}
\begin{indented}
\lineup
  \item[]\begin{tabular}{c|c|c|c|c}
  \br
         & \multicolumn{2}{c}{PAW} & \multicolumn{2}{c}{USPP} \\
    \hline
atom & valence & $r_{\mathrm{cut}}$ & valence & $r_{\mathrm{cut}}$ \\
    \hline
Fe & 3s 3p 3d 4s 4p & 2.0 (2.1$^d$) & 3d 4s 4p & 2.2 (2.3$^d$)\\
Ba & 5s 5p 5d & 2.9 & 5s 5p 6s & 2.5 (2.1$^p$)\\ 
As & 4s 4p 3d & 2.1 & 4s 4p & 2.4 (2.1$^p$)\\
  \br
  \end{tabular}
\end{indented}
\end{table}

\section{Results and Discussion}
 \subsection{\label{sec:esp}Electronic Structure Properties of Bulk BaFe$_2$As$_2$}

\begin{table}
  \caption{Comparison between our PAW-LDA results with FLAPW \cite{singh_prb_78_094511} and our USPP results with the same structure used in Ref. \cite{singh_prb_78_094511}. $\Delta E=E_{NM}-E_{tot}$, where $E_{tot}$ is the total energy of the specific spin-state and $E_{NM}$ is the NM state energy. The numbers in the parentheses are the same structure calculated using PBE exchange-correlation functional.}
  \label{TAB_FLAPW_PAW_CMP}
\begin{indented}
\lineup
  \item[]\begin{tabular}{c|cc|cc|}
  \br
   & \multicolumn{2}{c}{SDW-AFM} & \multicolumn{2}{c}{CB-AFM} \\
   \hline
   & {\it unrelaxed} & {\it opt-$z_{\mathrm{As}}$}
   & {\it unrelaxed} & {\it opt-$z_{\mathrm{As}}$} \\
   \hline
   $m_{\mathrm{Fe}}$ ($\mu_B$)& & & & \\
   FLAPW		& 1.75		& 0.70 
 				& 1.60		& N/A \\
   PAW			& 1.75(2.34) 		& 0.60(1.60)
  				& 1.54 		& N/A \\
   USPP			& 2.12		& 1.04 
   				& 1.90		& 0.58\\
   \hline
   $\Delta E$ (meV/Fe)& & & & \\
   FLAPW		& 92		& N/R 
 				& 41		& N/A \\
   PAW			& 65(174)		& $<1$(28) 
 				& 19		& N/A\\
   USPP			& 119		& 6.9	
 				& 60		& $<1$\\
 \br
  \end{tabular}
\end{indented}
\end{table}

\begin{figure}[htp]
  \centering
    \rotatebox{270}{\scalebox{0.7}{\includegraphics{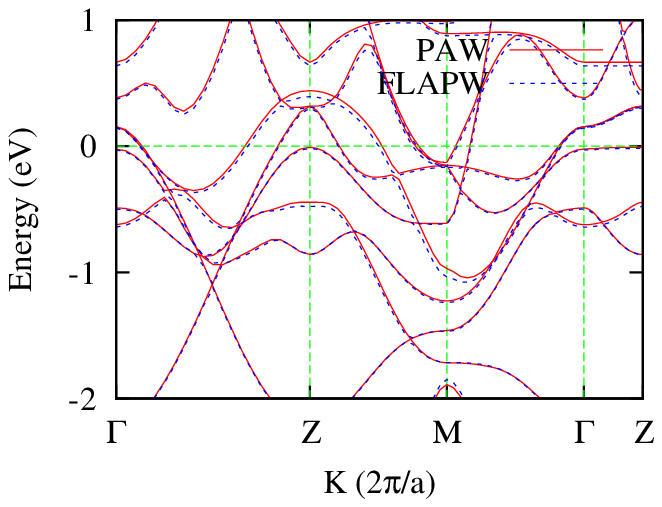}}}
  \caption{Comparison between FLAPW band structure and our PAW result for the non-magnetic state Ba122 with relaxed $z_{\mathrm{As}}$ but fixed experimental lattice parameters (LDA result). The PAW result (red solid line) is almost exactly the same as the FLAPW result (blue dashed line).\label{fig_bandstructure_relaxed_lda}}
\end{figure}

In order to directly compare with existing literatures using FLAPW method, we calculated the BaFe$_2$As$_2$ parental compound with different structural parameters. These calculations were performed on the structures used in Ref. \cite{singh_prb_78_094511}, using NM lattice parameters with relaxed $z_{\mathrm{As}}$ and with experimental $z_{\mathrm{As}}$. Both these parameters were used for the SDW-AFM state as well. Results are listed in Table \ref{TAB_FLAPW_PAW_CMP}. Our PAW calculations agree with FLAPW results very well, including the local magnetic moment on Fe atoms, whereas USPP always exagerates the magnetic moment, as reported previously. Using the LDA version of these datasets, we also calculated the band structure for the NM configuration relaxed-$z_{\mathrm{As}}$ structure. We compare our result and the FLAPW result in Fig. \ref{fig_bandstructure_relaxed_lda}, and it is apparent from this plot that our datasets almost exactly reproduced the FLAPW band structure. From Table \ref{TAB_FLAPW_PAW_CMP}, it is also observed that a relaxation of $z_{\mathrm{As}}$ will eventually bring the local magnetic moment down to $0.6 \mu_B$/Fe. In this situation, the SDW-AFM state energy is lower than the NM state energy by 65 meV/Fe. Neither FLAPW nor PAW is able to obtain a CB-AFM configuration with the relaxed $z_{\mathrm{As}}$, but this configuration remains stable using USPP. 

\begin{table}
  \caption{Physical properties after different levels of structural optimization using PAW. $z_{\mathrm{As}}$ refers to optimized $z_{\mathrm{As}}$ with experimental lattice constants, and {\it full-opt} means fully optimized structure with relaxed lattice constants. The values outside(inside) the parentheses are obtained with LDA(PBE). The SDW-AFM state does not exist after full relaxation using LDA, thus only PBE results are provided.}
  \label{TAB_Ba122_results}
  \begin{indented}
  \lineup
  \item[]\begin{tabular}{c|cc|cc}
  \br
   & \multicolumn{2}{c}{NM} & \multicolumn{2}{c}{SDW-AFM} \\
   \hline
   & {\it opt-$z_{\mathrm{As}}$} & {\it full-opt}
   & {\it opt-$z_{\mathrm{As}}$} & {\it full-opt} \\
   \hline
   $a$	& 3.963 & 3.873(3.968)& 5.615 &(5.697) \\
   $b$	& 3.963 & 3.873(3.968)& 5.574 &(5.594) \\
   $c$   & 13.02 & 12.14(12.49)& 12.94 &(12.76) \\
   $z_{\mathrm{As}}$ 					& 0.341(0.344) & 0.347(0.346) & 0.341(0.352) &(0.3514) \\
   $m_{\mathrm{Fe}}$ ($\mu_B$)	& 0.0 & 0.0 & 1.62(2.14)		&(2.19) \\
   $\Delta E$ (meV)					& 0.0 & 0.0 & 105 (83)		&(68)   \\
  \br
  \end{tabular}
\end{indented}
\end{table}

\begin{figure}[htp]
  \centering
  \subfigure[{\it unrelaxed}] {
    \label{fig_bandstructure_expt}
    \rotatebox{270}{\scalebox{0.5}{\includegraphics{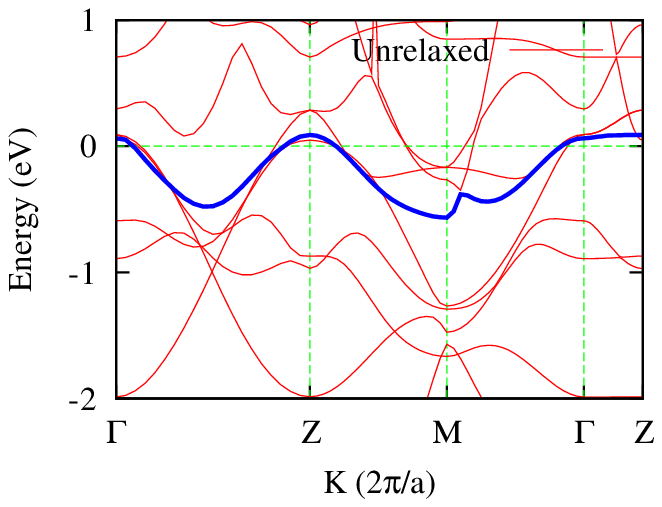}}}
  }
  \subfigure[{\it full-opt}] {
    \label{fig_bandstructure_vcrelax}
    \rotatebox{270}{\scalebox{0.5}{\includegraphics{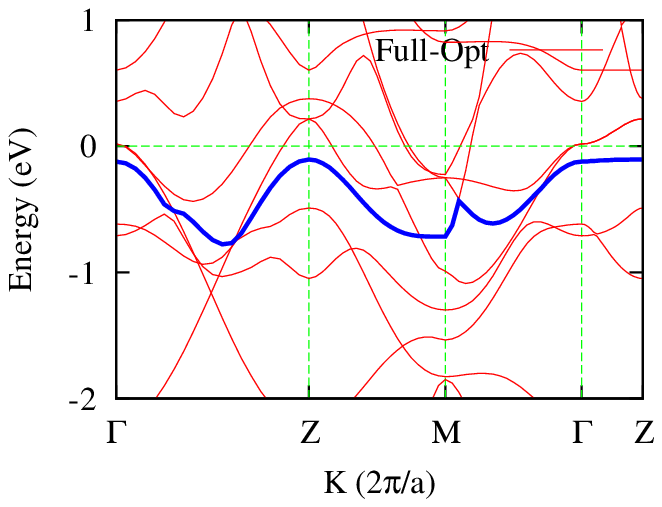}}}
  }
  \caption{Band structure for NM state Ba122 with (a) experimental structure, and (b) fully optimized lattice parameters as well as $z_{\mathrm{As}}$ using PBE exchange-correlation functional. The band structure for the {\it opt-$z_{\mathrm{As}}$} structure looks very close to the LDA result (Fig. \ref{fig_bandstructure_relaxed_lda}), and is therefore not shown here. 
The effect of structural optimization mostly appears in the thick blue band.
  \label{fig_bandstructure}}
\end{figure}

\begin{figure}[htp]
  \centering
  \subfigure[{\it unrelaxed}] {
    \label{fig_sdw_bs_expt}
    \rotatebox{270}{\scalebox{0.5}{\includegraphics{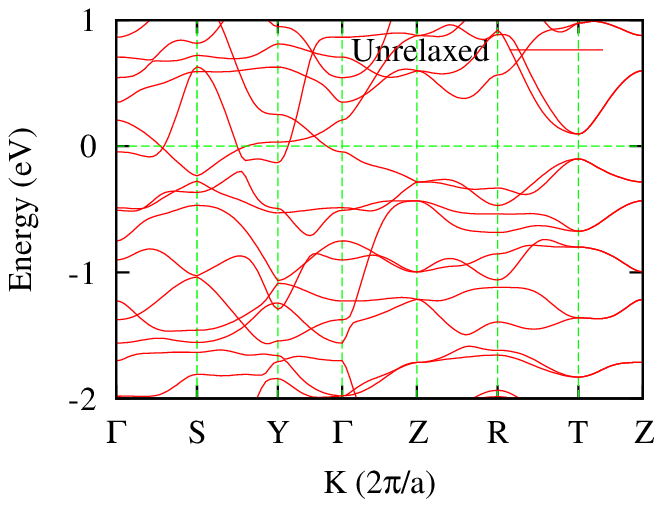}}}
  }
  \subfigure[{\it full-opt}] {
    \label{fig_sdw_bs_vcrelax}
    \rotatebox{270}{\scalebox{0.5}{\includegraphics{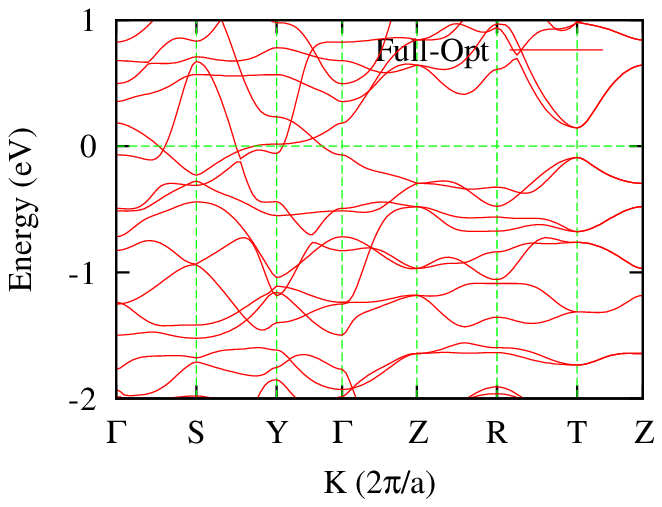}}}
  }
  \caption{Band structure for SDW-AFM state Ba122 with (a) experimental structure, and (b) fully optimized lattice parameters as well as $z_{\mathrm{As}}$ using PBE exchange-correlation functional. 
Due to the spin degeneracy, only the $\alpha$ spin band structure is shown here.\label{fig_sdw_bs}}
\end{figure}

Next, to evaluate different exchange-correlation functionals we study the band structure dependency on the degree structural optimization. As summarized in Table \ref{TAB_Ba122_results}, the PBE exchange-correlation functional always exagerates the local magnetic moments on Fe atoms, and the magnetic moment remains relatively unchanged with respect to the structural optimizations. However, the PBE exchange-correlation functional yields crystal structures much closer to experiment, and the $z_{\mathrm{As}}$ parameters are significantly different using PBE and using LDA. PBE also yields a large magnetic moment change when the partially-optimized LDA structure ($z_{\mathrm{As}}$ optimized with LDA but lattice constants fixed at experimental values) is used; see Table \ref{TAB_FLAPW_PAW_CMP} data in parentheses. It was also observed that a full optimization of both lattice parameters and $z_{\mathrm{As}}$ within LDA fails to yield a SDW-AFM ground state (in fact, it does not show any magnetic instability after full optimization) due to the fact that LDA underestimates bond lengths and lattice constants while this system is too delicate to allow such underestimation. Mazin {\it et al.} have observed similar results, that LDA will sometimes miss the SDW-AFM ground state. PBE overestimates both bond lengths and lattice constants, obtaining the correct ground state of the system under full optimization due to an error cancellation. Both exchange-correlation functionals underestimate the $c$-axis in both the NM and SDW-AFM states, indicating an  incompleteness of LDA and GGA exchange-correlation functionals. Since the local magnetic moment of Fe atoms depends strongly on $z_{\mathrm{As}}$ parameters, as shown in previous studies and in this work, we conclude that for the magnetic properties, neither LDA nor PBE is capable of describing the subtleness of this particular system; but PBE correctly describes the ground state of the system under full optimization.

Due to the loss of the SDW-AFM ground state in LDA after full optimization, the following discussion includes PBE  results only. As pointed out by Refs. \cite{dft_fix_z_1} and \cite{dft_fix_z_2}, DFT calculations can be comparable with experimental results only if $z_{\mathrm{As}}$ is fixed at the experimental value. Thus, these calculations were performed with FLAPW without full optimization of lattice parameters. Figure \ref{fig_bandstructure} depicts a set of band structures corresponding to different levels of structural optimization. Apparently, optimization does not alter significantly the electronic structure around the M point, where two electron pockets were found. Moreover, even the size of these electron pockets does not strongly depend on the structural optimization. In contrast, the structural relaxation substantially changed the band structure around $\Gamma$ and Z. The main effect of structural relaxation is that optimization pushes one of the Fe $3d$-bands toward lower energy (the thick blue solid line in Fig. \ref{fig_bandstructure}); thus, only the two hole pockets from the rest of the $3d$-bands remains after $z_{\mathrm{As}}$ optimization as well as after full structural relaxation. 

Unlike the NM state, the band structure of SDW-AFM state does not exhibit large variation before and after structural optimization (Fig. \ref{fig_sdw_bs}). This can be easily understood since $z_{\mathrm{As}}$ in SDW-AFM state was maintained during structure relaxation. We have used the SDW-AFM experimental structure and calculated the band structure with LDA, and there is not much difference between the LDA result and the ones shown in Fig \ref{fig_sdw_bs}, despite of the significant difference ($ 0.5 \mu_B$) in the $m_{\mathrm{Fe}}$ found in these calculations. It is then apparent that the fundamental discrepancy between the LDA and the GGA calculations in the previous literature is due to the difference between the structure optimized with LDA and GGA, respectively. The full optimization with the PBE exchange-correlation functional does not significantly alter the band structure in the SDW-AFM state, but is crucial to study realistic doping and pressure effects in order to be consistent.

\begin{figure}[htp]
  \centering
  \subfigure[NM-{\it unrelaxed}] {
    \label{fig_pm_fs_expt}
    \rotatebox{270}{\scalebox{0.5}{\includegraphics{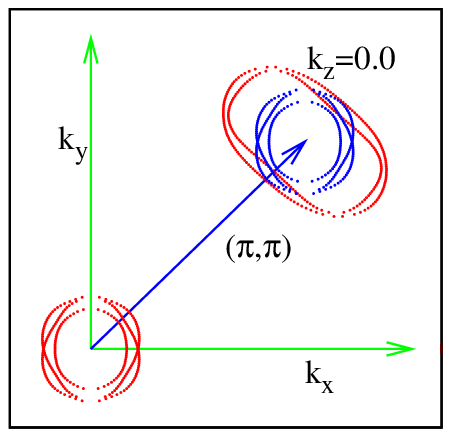}}}
  }
  \subfigure[NM-{\it full-opt}] {
    \label{fig_pm_fs_fullopt}
    \rotatebox{270}{\scalebox{0.5}{\includegraphics{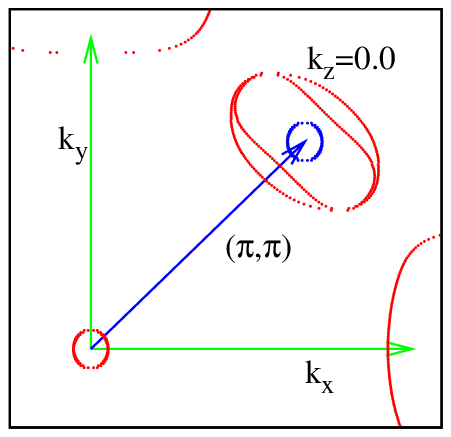}}}
  }
  \subfigure[SDW-{\it full-opt}] {
    \label{fig_sdw_fs_fullopt}
    \rotatebox{270}{\scalebox{0.5}{\includegraphics{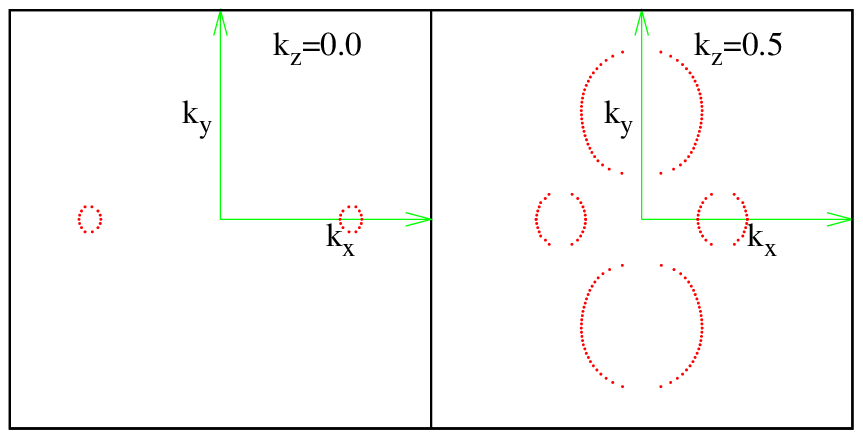}}}
  }
  \caption{Fermi surfaces reconstructed using maximally localized Wannier functions (MLWFs). $k_x$ and $k_y$ 
denote the reciprocal vectors corresponding to $a$ and $b$ for the conventional tetragonal unit cell.\label{fig_fs}}
\end{figure}

The effect of structural optimization on the electronic structure can also be illustrated with the Fermi surfaces. In Fig. \ref{fig_fs} we present the Fermi surfaces reconstructed using maximally localized Wannier functions\cite{MLWF_1,MLWF_2} (MLWFs). For the NM state, as we have seen in Fig. \ref{fig_bandstructure}, the optimization leads to very different band structures, and thus results in very different Fermi surfaces as well. After full-optimization, the size of hole pockets around $\Gamma$ is greatly reduced and the two electron pockets are much less affected, as suggested by the band structure calculation, and therefore the expected ($\pi$,$\pi$) Fermi-surface nesting is missing. For the SDW state, since the optimization does not significantly alter the electronic structure as it does in the NM state, we thus show only the Fermi surfaces after full optimization. Unlike its NM state, the SDW Fermi surfaces are surprisingly simple and do not show very interesting features such as Fermi-surface nesting. Two extended electron pockets were found, and two hole pockets appear around $k_z=0.5$.

\subsection{\label{sec:dfp}Diffusion of Ba Atom on BaFe$_2$As$_2$ (001) Surface}
The BaFe$_2$As$_2$ (001) surface can in principle be produced with two ways: 1. (type-I) separation between Fe-As layers to form Fe-terminated or As-terminated surfaces; 2. (type-II) separation between As-Ba layers to form As-terminated or Ba-terminated surfaces. However, type-I surfaces are prohibited energetically due to the strong Fe-As bonds. 

\begin{figure}[htp]
  \centering
    \subfigure[initial] {
    \label{fig_diffusion_initial}
    \scalebox{0.25}{\includegraphics{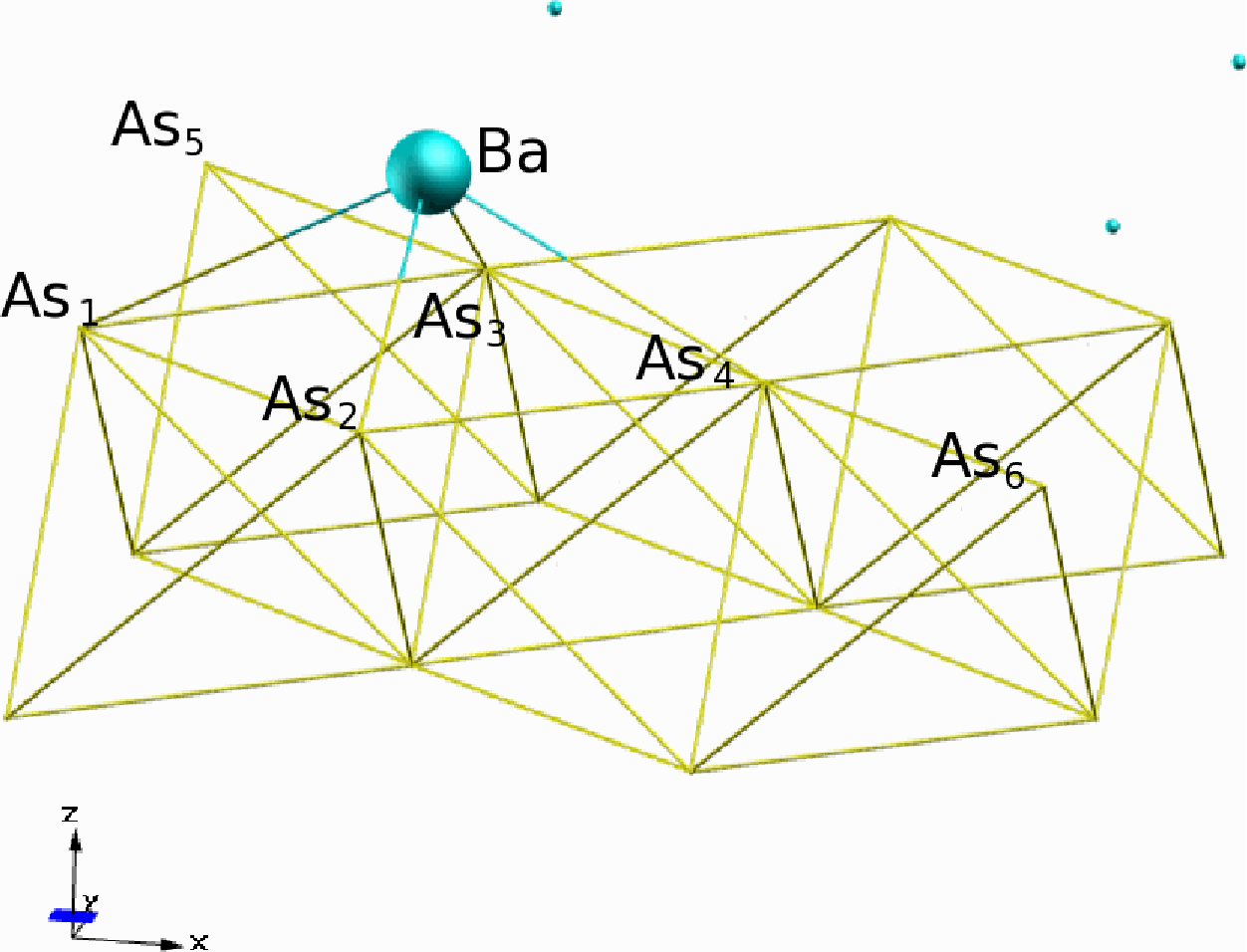}}
  }
  \subfigure[uphill 1] {
    \label{fig_diffusion_up1}
    \scalebox{0.25}{\includegraphics{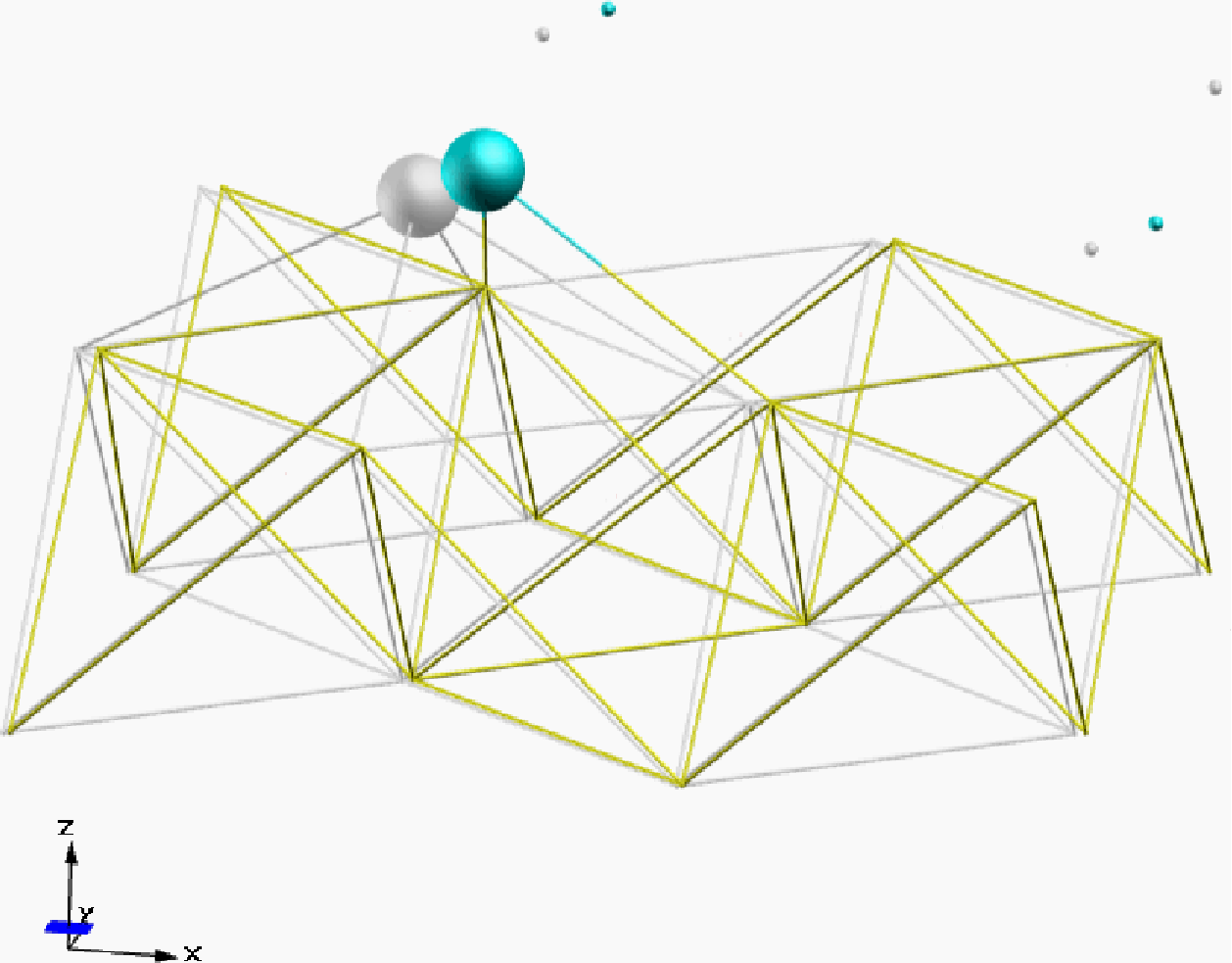}}
  }
  \subfigure[uphill 2] {
    \label{fig_diffusion_up2}
    \scalebox{0.25}{\includegraphics{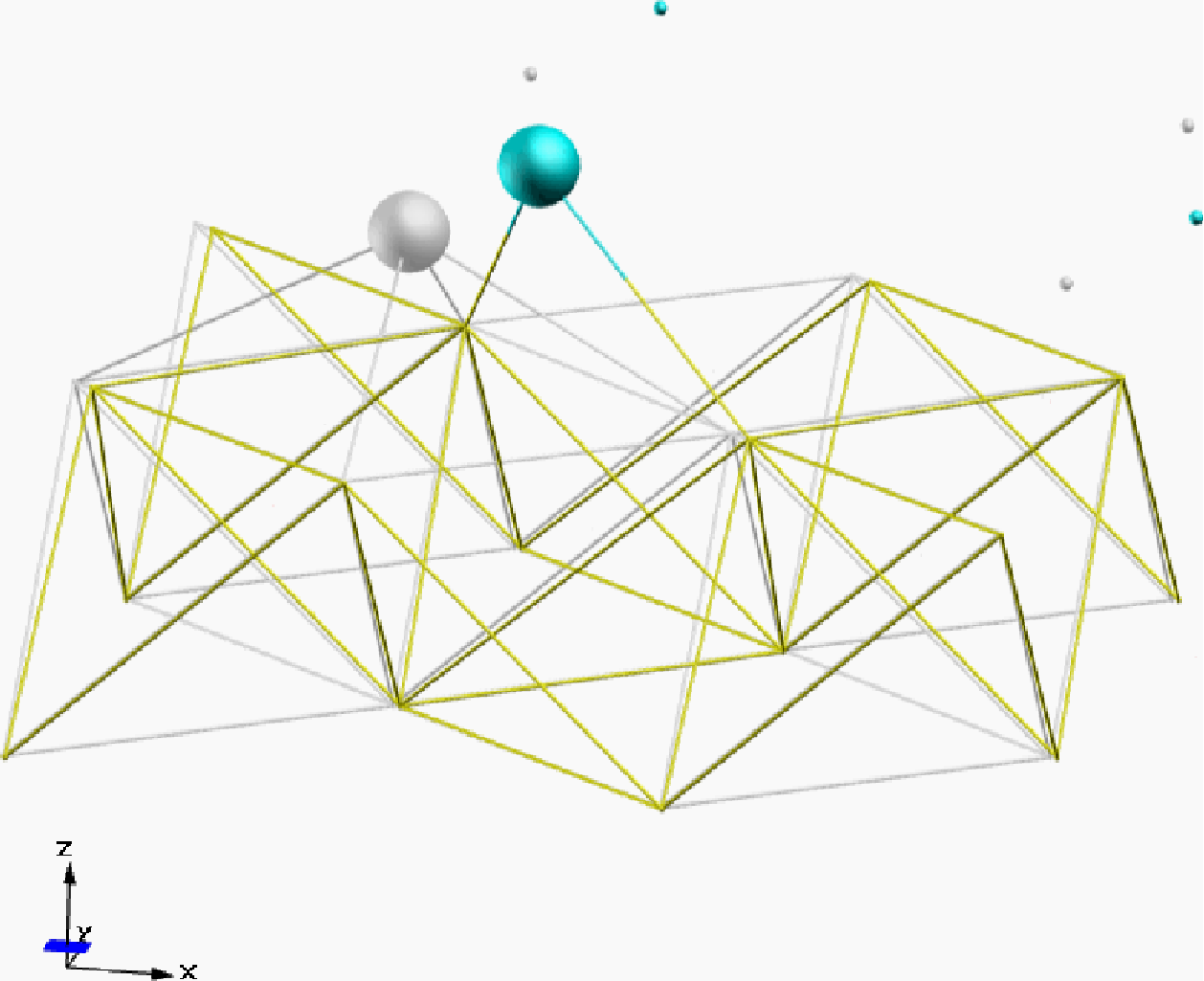}}
  }
  \subfigure[transition state] {
    \label{fig_diffusion_peak}
    \scalebox{0.25}{\includegraphics{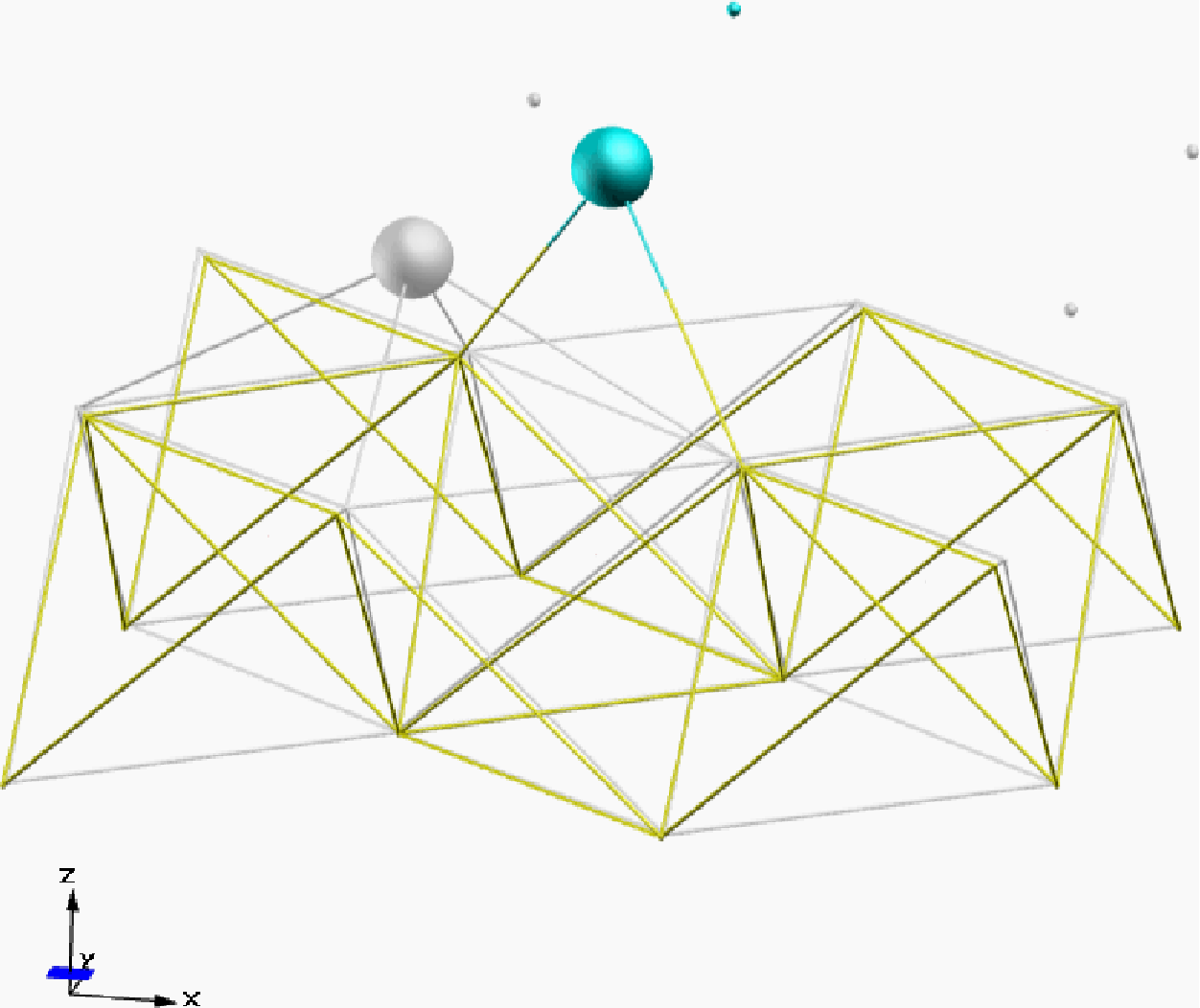}}
  }
  \caption{Diffusion of Ba atom on As-terminated type-II surface. The process is demonstrated using a 2$\times$2 unit cell used in the simulation, and since the uphill process is symmetric to the downhill process, only the uphill process is shown here. In order to present a clear visualization, we show only the surface Ba atoms and the topmost layer of the As structure, and exaggerate one of the Ba atoms on the surface as well. The Ba atoms are displayed with balls, and the As structure is shown with the As tetragon. For comparison purposes, figures in panels b), c), d) and e) are superposed on the initial structure (the grayed structure) as well. \label{fig_diffusion}}
\end{figure}

\begin{figure}[htp]
  \centering
  \rotatebox{270}{\scalebox{0.7}{\includegraphics{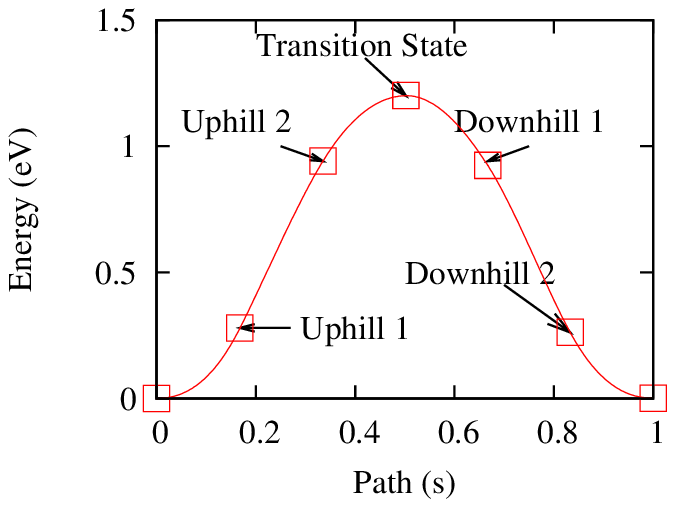}}}
  \caption{Ba atom diffusion energy barrier. $s$ is the reaction parameter, which is defined by $d_{Ba}/d_{Ba}^{0}$, where $d_{Ba}$ is the distance the Ba atom has traveled and $d_{Ba}^{0}$ is the distance between the neighbouring equilibrium positions for the Ba atom on the As-terminated type-II surface. $s=0$ and $s=1$ represents initial and final states, respectively.\label{fig_diffuse_barrier}}
\end{figure}

For type-II surfaces, since the Ba-layers are sandwiched by As-layers, both As-termination and Ba-termination are possible. STM+STS measurements on parental BaFe$_2$As$_2$ surface by Nascimento {\it et al.}\cite{PRL_Ba122_surface} shows $\sqrt{2}\times\sqrt{2}$ ordered structure and disordered bright spots, which they explained as disordered Ba atoms on ordered As-terminated surfaces, and that the ordered structure in STM is due to surface As atom. However, experiments on Co-doped material BaFe$_{2-x}$Co$_x$As$_2$\cite{PRB_Ba122_Co_surface,arxiv.1009.3493} shows not only the ordered $\sqrt{2}\times\sqrt{2}$ structure and the disordered structure, but ordered $2\times 1$ structures as well. They claimed that both ordered and disordered STM structures are due to surface Ba atoms and that the ordering is a result of surface reconstruction with increased temperature.
It is then interesting to study the diffusion path, transition state and diffusion energy barrier of Ba atom on an As-terminated type-II surface, and to identify the ordered structures. Therefore, we have performed a nudged elastic band (NEB)\cite{NEB_1,NEB_2} calculation for this system. The starting and final configurations were taken to be two neighbouring equilibrium lowest energy states, where the Ba atom was located above the center of four surface As atoms (Fig. \ref{fig_diffusion_initial}). Seven NEB images, including the starting and final images, were considered within the calculation, and the convergence criterion for the normal forces were chosen to be 0.04 eV/\AA. The result shows that the Ba atoms diffuse along the diagonal of the orthorhombic lattice, and the energy barrier turns out to be 1.2 eV (Fig. \ref{fig_diffuse_barrier}). 

A close examination of the intermediate states reveals more details about the diffusion process, as shown in Fig. \ref{fig_diffusion}. At equilibrium, four As atoms form chemical bonds with the Ba atom (Fig. \ref{fig_diffusion_initial}). We denote these four atoms with As$_1$ to As$_4$,  and two more As atoms with As$_5$ and As$_6$ as shown in Fig. \ref{fig_diffusion_initial}. At equilibrium, the distances between all nearest neighbour As atoms in the topmost As layer are, e.g. $d_{As_1-As_2}=d_{As_2-As_3}=d_{As_3-As_4}=d_{As_4-As_1}=d_{As_3-As_5}=d_{As_4-As_6}=3.99$ \AA. However, the topmost As layer does not form a square lattice, since the Ba-As bonds can be classified into two categories: $d_{Ba-As_2}=d_{Ba-As_3}=3.18$ \AA\ and $d_{Ba-As_1}=d_{Ba-As_4}=3.24$\AA. The Ba atom then diffuse along the (110) direction. At snapshot "uphill 1" (Fig.\ref{fig_diffusion_up1}), the topmost As layer shift towards the (110) direction as well, causing the distortion of the As tetragons. The distance between the As$_3$ and As$_4$ atoms is stretched to 4.15 \AA, and $d_{As_1-As_2}$ is reduced to 3.83 \AA. The Ba-As$_1$ and Ba-As$_2$ bonds breaks, while the $d_{Ba-As_3}$ and $d_{Ba-As_4}$ bonds are reduced to 3.07 \AA\ and 3.09 \AA, respectively. At snapshot "uphill 2" (Fig.\ref{fig_diffusion_up2}), the topmost As layer slightly moves backwards along the ($\overline{11}$0) direction due to the strain caused by the As tetragon distortion, and $d_{As_1-As_2}$ and $d_{As_3-As_4}$ are now 3.79 \AA and 4.20 \AA, respectively. The Ba-As$_3$ and Ba-As$_4$ bonds are further reduced to 3.01 \AA. At the transition state (Fig.\ref{fig_diffusion_peak}), the Ba-As$_3$, Ba-As$_4$, As$_1$-As$_2$, and As$_3$-As$_4$ bond lengths remain unchanged from snapshot "uphill 2", but the As$_3$, As$_4$ and Ba atoms are now inside the same [110] plane. It is obvious that during this process, two major structural changes occur simultaneously, the distortion of the As tetragons (and thus the internal stress within the Fe-As layers) and Ba-As bond breaking. The diffusion energy barrier is mainly due to the Ba-As bond breaking process, since no obvious As-tetragon change can be observed from snapshot "uphill 2" to the transition state. Our result indicates that the Ba atom clearly has a favorable adsorption site on the As-terminated type-II surface, although no long-range order is suggested.

\begin{figure}[htp]
  \centering
  \subfigure[As-terminated] {
    \label{fig_STM_surface_as}
    \scalebox{0.3}{\includegraphics{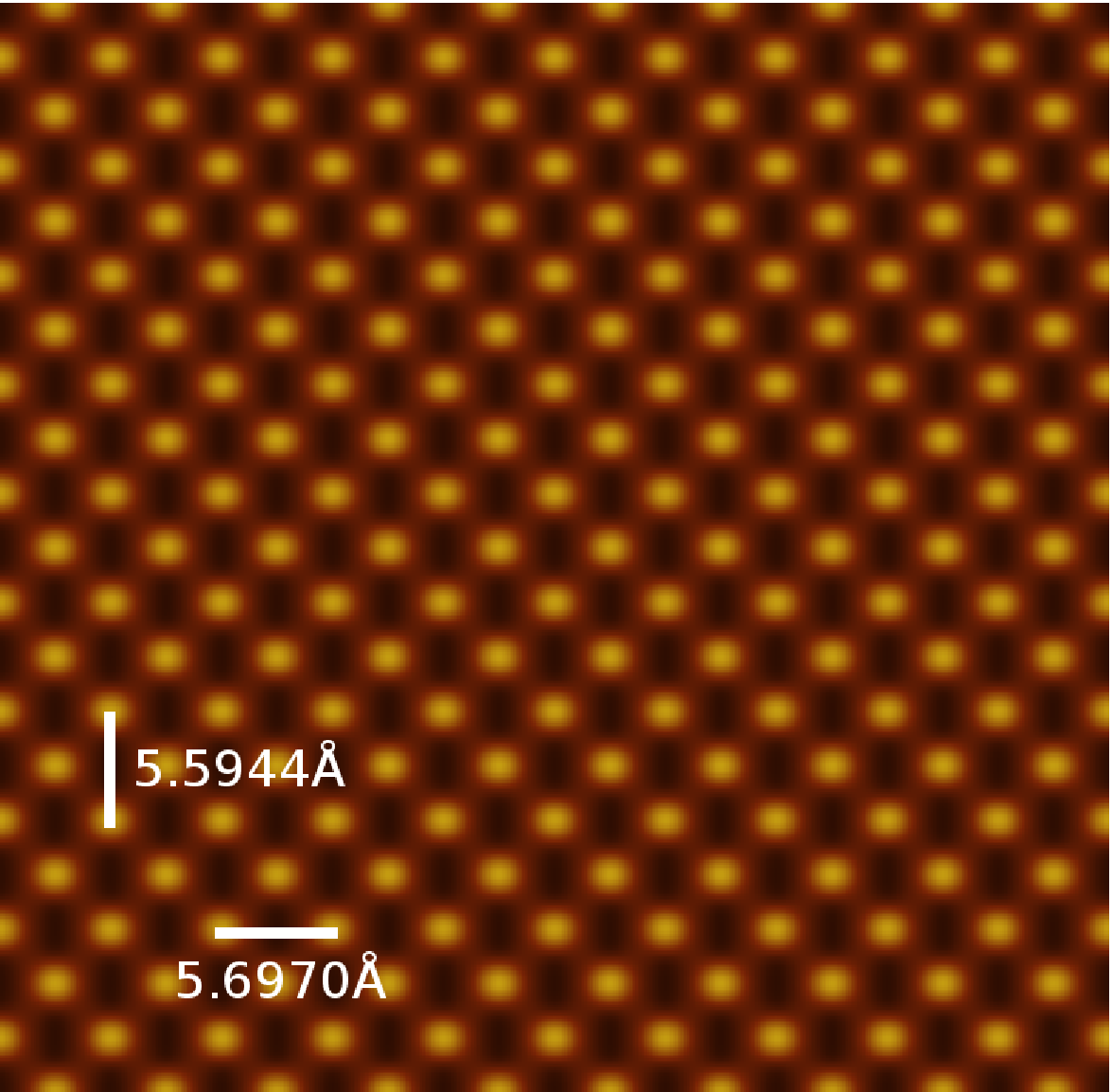}}
  }
  \subfigure[Ba-terminated] {
    \label{fig_STM_surface_ba}
    \scalebox{0.3}{\includegraphics{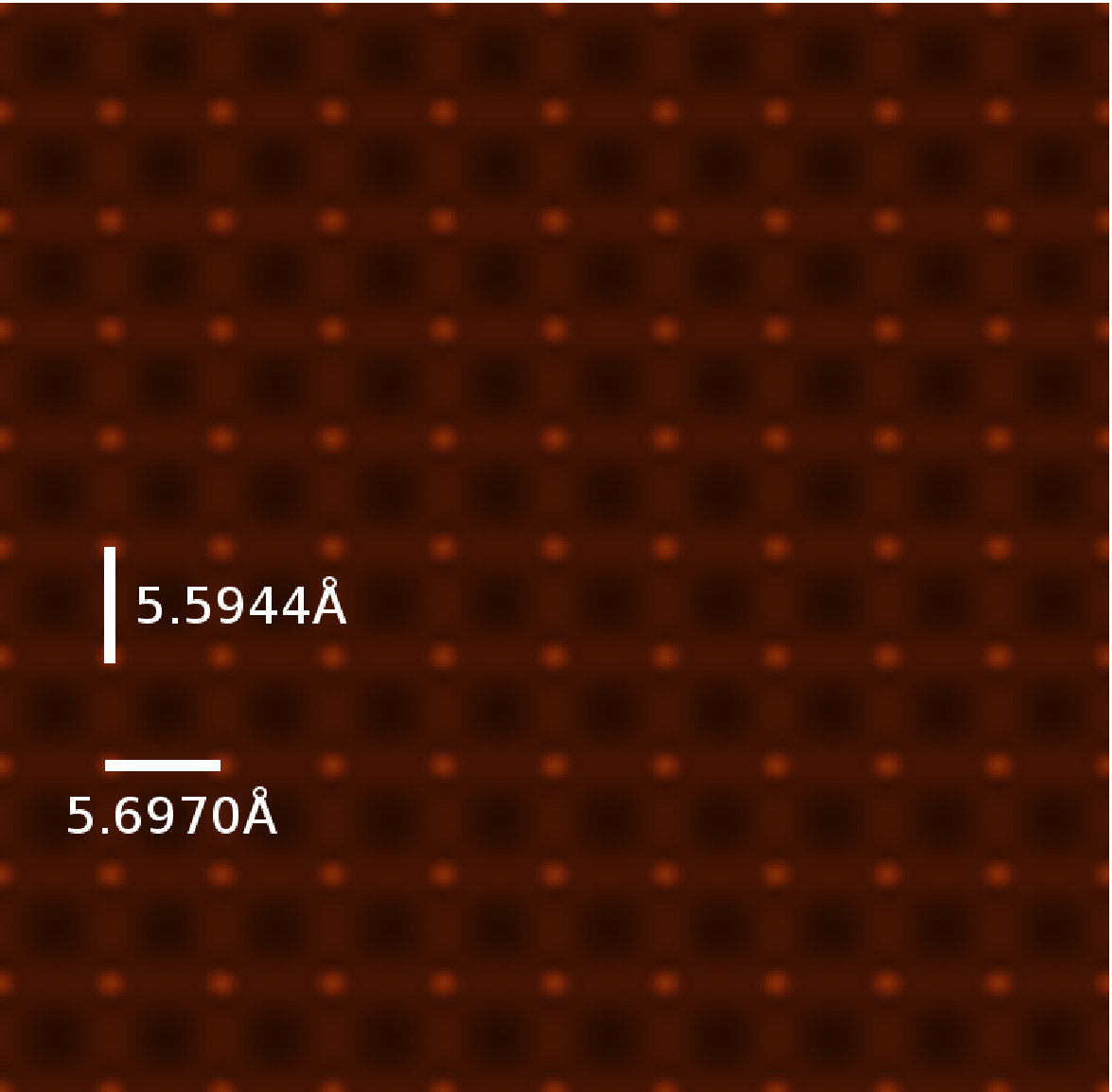}}
  }
  \caption{Simulated STM image of BaFe$_2$As$_2$ surface. The figures show a $10\times10$ simulated surface structure. Simulated bias voltage $V_b$=-0.02 \Ry.\label{fig_STM_surface}}
\end{figure}

In order to compare our results with the experiments, we have also performed local density of states (LDOS) calculations to simulate the STM image. As discussed earlier, both As-terminated and Ba-terminated type II surfaces are possible, and we thus simulated both cases (FIG. \ref{fig_STM_surface}). Nascimento {\it et al.}\cite{PRL_Ba122_surface} suggest that only one surface As atom out of two per unit cell can be observed by STM due to an in-plane orthorhombic distortion of the Fe atoms, and that the surface is covered by randomly displaced Ba atoms. However, although our surface has the proposed orthorhombic distortion with the orthorhombicity $\epsilon=1-b/a$ calculated to be 1.8\%, we were unable to obtain a similar STM image; unless the STM image records the position of Ba atoms and the Ba atoms form a $\sqrt{2}\times\sqrt{2}$ lattice over the surface (FIG. \ref{fig_STM_surface_ba}). Our result is in consistency with F. Massee {\it et al.}, although we did not perform calculations with $2\times 1$ ordered structure since this feature is suggested to be a consequence of Co-doping and does not appear in prestine BaFe$_2$As$_2$. The As-terminated surface always shows both surface As atoms (FIG. \ref{fig_STM_surface_as}). To verify this result, we have explored a wide $V_b$ range from -0.1 \Ry\ to 0.1 \Ry, while the experiment was conducted at $V_b=-0.1 V$ and $-0.02 V$. Our result indicates that: 1 the Ba atom clearly has a favorable adsorption site on the As-terminated type-II surface; 2 the STM image shows LDOS from Ba instead of As atoms; and 3 the ordered $\sqrt{2}\times\sqrt{2}$ STM image indicates an ordered structure of surface Ba atoms

\section{Conclusion}
In conclusion, we have constructed a reliable PAW dataset that yields results comparable to the FLAPW method, that we then used to study electronic and magnetic structure for BaFe$_2$As$_2$. A thorough comparison between exchange-correlation functionals as well as structural optimizations were presented. It is found that the LDA exchange-correlation functional misses the SDW-AFM configuration under full structural optimization, and the PBE gives the correct ground state. The important $c$-axis lattice parameter and $z_{\mathrm{As}}$ can be obtained within reasonable error for the SDW-AFM state with the PBE exchange-correlation functional, but both values are noticeably reduced for the NM state. Therefore, the electronic structure of the SDW-AFM state is insensitive to structural optimization, but the optimization will yield unacceptable electronic structure for the NM state. PBE can describe the SDW-AFM ground state very well but not the NM state, and the LDA fails for both. The apparent correlation between magnetism and crystal structure suggests that LDA/GGA fails for the NM state because of a strong magnetic-phonon coupling within the iron-pnictide compounds, and that this strong coupling may be the mechanism for the electron pairing. We have also simulated the surface properties of BaFe$_2$As$_2$. The simulated STM image is inconsistent with the experiment, which is probably due to insufficient experimental resolution. It was also found that the Ba atom diffuses on the As-terminated BaFe$_2$As$_2$ with a large barrier of 1.2 eV, and that the Ba atom has a well-defined adsorption site on this surface. The barrier suggests a low mobility of the Ba atoms on the surface.

This work is supported by NSF-DMR. The authors would like to thank the UF-HPC center, ORNL CNMS user program, and DOE/NERSC for the computational resources that have contributed to the research results reported within this paper. The PAW datasets and the parameters used to generate them are available on http://www.qtp.ufl.edu/igator/PAW.htm.

\section*{References}
\bibliographystyle{unsrt.bst}
\bibliography{paw_gen}

\begin{thebibliography}{10}

\bibitem{la1111_discover}
Y.~Kamihara, T.~Watanabe, M.~Hirano, and H.~Hosono.
\newblock {\em J.\ Am.\ Chem.\ Soc.}, 130:3296, 2008.

\bibitem{ba122_discover}
Marianne Rotter, Marcus Tegel, and Dirk Johrendt.
\newblock {\em Phys.\ Rev.\ B}, 78:020503, 2008.

\bibitem{Li111_discover_1}
M.~J. Pitcher, D.~R. Parker, P.~Adamson, S.~J.~C. Herkelrath, A.~T. Boothroyd,
  R.~M. Ibberson, M.~Brunelli, and S.~J. Clarke.
\newblock {\em Chem. Commun.}, page 5918, 2008.

\bibitem{Li111_discover_2}
Joshua~H. Tapp, Zhongjia Tang, Bing Lv, Kalyan Sasmal, Bernd Lorenz, Paul C.~W.
  Chu, and Arnold~M. Guloy1.
\newblock {\em Phys. Rev. B}, 78:060505(R), 2008.

\bibitem{Nature.453.899}
Clarina de~la Cruz, Q.~Huang, J.~W. Lynn, Jiying Li, W.~Ratcliff II, J.~L.
  Zarestky, H.~A. Mook, G.~F. Chen, J.~L. Luo, N.~L. Wang, and Pengcheng Dai.
\newblock {\em Nature}, 453:899--902, 2008.

\bibitem{PhysRevB.78.020503}
Marianne Rotter, Marcus Tegel, Dirk Johrendt, Inga Schellenberg, Wilfried
  Hermes, and Rainer P\"ottgen.
\newblock Spin-density-wave anomaly at 140 k in the ternary iron arsenide $
  bafe2 as2 $.
\newblock {\em Phys. Rev. B}, 78(2):020503, Jul 2008.

\bibitem{PhysRevB.78.094517}
Michael~A. McGuire, Andrew~D. Christianson, Athena~S. Sefat, Brian~C. Sales,
  Mark~D. Lumsden, Rongying Jin, E.~Andrew Payzant, David Mandrus, Yanbing
  Luan, Veerle Keppens, Vijayalaksmi Varadarajan, Joseph~W. Brill, Rapha\"el~P.
  Hermann, Moulay~T. Sougrati, Fernande Grandjean, and Gary~J. Long.
\newblock Phase transitions in lafeaso: Structural, magnetic, elastic, and
  transport properties, heat capacity and m\"ossbauer spectra.
\newblock {\em Phys. Rev. B}, 78(9):094517, Sep 2008.

\bibitem{singh_prl_100_237003}
D.~J. Singh and M.-H. Du.
\newblock {\em Phys.\ Rev.\ Lett.}, 100:237003, 2008.

\bibitem{ccao_1}
Chao Cao, P.~J. Hirschfeld, and Hai-Ping Cheng.
\newblock {\em Phys.\ Rev.\ B}, 77:220506, 2008.

\bibitem{yildirim_1}
T.~Yildirim.
\newblock {\em Phys.\ Rev.\ Lett.}, 101:057010, 2008.

\bibitem{yildirim_2}
T.~Yildirim.
\newblock {\em Phys.\ Rev.\ Lett.}, 102:037003, 2009.

\bibitem{vanderbilt_uspp}
David Vanderbilt.
\newblock {\em Phys.\ Rev.\ B}, 41:7892, 1990.

\bibitem{bloch_paw}
P.~E. Bloch.
\newblock {\em Phys. Rev. B}, 50:17953, 1994.

\bibitem{flapw}
D.J. Singh and L.~Nordstrom.
\newblock {\em Planewaves, Pseudopotentials and the LAPW Method, 2nd. Ed.}
\newblock Springer, Berlin, 2006.

\bibitem{PhysRevB.78.085104}
I.~I. Mazin, M.~D. Johannes, L.~Boeri, K.~Koepernik, and D.~J. Singh.
\newblock Problems with reconciling density functional theory calculations with
  experiment in ferropnictides.
\newblock {\em Phys. Rev. B}, 78(8):085104, Aug 2008.

\bibitem{PWSCF}
{\em PWSCF in Quantum Espresso Package}, 2007.

\bibitem{vasp_paw}
G.~Kresse and D.~Joubert.
\newblock {\em Phys.\ Rev.\ B}, 59:1758, 1999.

\bibitem{mp_kpoints}
H.~J. Monkhorst and J.~D. Pack.
\newblock {\em Phys. Rev. B}, 13:5188, 1976.

\bibitem{PBE_1}
J.~P. Perdew, K.~Burke, and M.~Ernzerhof.
\newblock {\em Phys. Rev. Lett.}, 77:3865, 1996.

\bibitem{singh_prb_78_094511}
D.~J. Singh.
\newblock {\em Phys.\ Rev.\ B}, 78:094511, 2008.

\bibitem{dft_fix_z_1}
Deepa Kasinathan, Alim Ormeci, Katrin Koch, Ulrich Burkhardt, Walter Schnelle,
  Andreas Leithe-Jasper, and Helge Rosner.
\newblock {\em New J. Phys.}, 11:025023, 2009.

\bibitem{dft_fix_z_2}
S.~A.~J. Kimber, A.~Kreyssig, Y.Z. Zhang, H.O. Jeschke, R.~Valent\'{i},
  F.~Yokaichiya, E.~Colombier, J.-Q. Yan, T.~C. Hansen, T.~Chatterji, R.J.
  McQueeney, P.C. Canfield, A.I. Goldman, and D.~N. Argyriou, 2009.
\newblock in press.

\bibitem{MLWF_1}
I.~Souza, N.~Marzari, and D.~Vanderbilt.
\newblock {\em Phys. Rev. B}, 65:035109, 2001.

\bibitem{MLWF_2}
A.~A. Mostofi, J.~R. Yates, Y.-S. Lee, I.~Souza, D.~Vanderbilt, and N.~Marzari.
\newblock {\em Comp. Phys. Comm.}, 178:685, 2008.

\bibitem{PRL_Ba122_surface}
V.~B. Nascimento, Ang Li, Dilushan~R. Jayasundara, Yi~Xuan, Jared O'Neal,
  Shuheng Pan, T.~Y. Chen, Biao Hu, X.~B. He, Guorong Li, A.~S. Sefat, M.~A.
  McGuire, B.~C. Sales, D.~Mandrus, M.~H. Pan, Jiandi Zhang, R.~Jin, and E.~W.
  Plummer.
\newblock {\em Phys. Rev. Lett.}, 103:076104, 2009.

\bibitem{PRB_Ba122_Co_surface}
F.~Massee, S.~de~Jong, Y.~Huang, J.~Kaas, E.~van Heumen, J.~B. Goedkoop, and
  M.~S. Golden.
\newblock {\em Phys. Rev. B}, 80:140507(R), 2009.

\bibitem{arxiv.1009.3493}
Erik van Heumen, Johannes Vuorinen, Klaus Koepernik, Freek Massee, Yingkai
  Huang, Ming Shi, Jesse Klei, Jeroen Goedkoop, Matti Lindroos, Jeroen van~den
  Brink, and Mark~S. Golden.
\newblock {\em arxiv: eprints}, page 1009.3493v1, 2010.

\bibitem{NEB_1}
G.~Mills and H.~J\'{o}nsson.
\newblock {\em Phys. Rev. Lett.}, 72(1124), 1994.

\bibitem{NEB_2}
G.~Mills, H.~J\'{o}sson, and G.~K. Schenter.
\newblock {\em Surf. Sci.}, 324(305), 1995.

\end{thebibliography}
\end{document}